\documentclass[prb,showpacs,amsmath,amssymb,superscriptaddress,twocolumn,floatfix]{revtex4-1}

\usepackage{graphicx}
\usepackage{color}

\begin{document}

\title{Anomalous Spectroscopical Effects in an Antiferromagnetic Semiconductor}

\author{M. Hubert}
\affiliation{Faculty of Mathematics and Physics, 
  Charles University in Prague, Praha, CZ-121 16, Czech Republic}

\author{T. Male\v cek}
\affiliation{Faculty of Mathematics and Physics, 
  Charles University in Prague, Praha, CZ-121 16, Czech Republic}

\author{K.--H. Ahn}
\affiliation{Institute of Physics, ASCR, $v.~v.~i.$, 
Cukrovarnick\'a 10, CZ-16253 Praha 6, Czech Republic}

\author{M. M\'\i{}\v sek}
\affiliation{Institute of Physics, ASCR, $v.~v.~i.$, 
Cukrovarnick\'a 10, CZ-16253 Praha 6, Czech Republic}

\author{J.~\v Zelezn\'y}
\affiliation{Institute of Physics, ASCR, $v.~v.~i.$, 
Cukrovarnick\'a 10, CZ-16253 Praha 6, Czech Republic}

\author{F. M\'aca}
\affiliation{Institute of Physics, ASCR, $v.~v.~i.$, 
Na Slovance 2, CZ-18221 Praha 8, Czech Republic}

\author{G. Springholz}
\affiliation{Institute of Semicond. Sol. State Physics, Johannes Kepler Univ. Linz, Altenbergerstr. 69, A-4040 Linz, Austria}

\author{M. Veis}
\affiliation{Faculty of Mathematics and Physics, 
  Charles University in Prague, Praha, CZ-121 16, Czech Republic}

\author{K. V\'yborn\'y}
\affiliation{Institute of Physics, ASCR, $v.~v.~i.$, 
Cukrovarnick\'a 10, CZ-16253 Praha 6, Czech Republic}

\date{Nov18, 2024}

\begin{abstract}
Following the recent observation of anomalous Hall effect in
antiferromagnetic hexagonal MnTe thin  
films, related phenomena at finite frequencies have come into focus. Magnetic
circular dichroism (MCD) is the key material property here. In the x-ray
range, the XMCD has already been demonstrated and used to visualise domains
via photoemission electron microscopy (PEEM). Here we report on MCD in
optical range and discuss its microscopic mechanism.

\end{abstract}


\maketitle

\section{Introduction}

Several aspects of manganese telluride~\cite{Allen:1977_a} are worth
highlighting: its structure in relation to similar compounds, the semiconducting
character and also, the magnetic order it supports. Regarding the last one, 
the anomalous Hall effect~\cite{Nagaosa:2010_a} (AHE) has long been 
associated only with ferromagnets and often (yet not 
always\cite{Mathieu:2004_a,Zelezny:2023_a})
it was assumed to be proportional to magnetisation. Occasional observations
which proved these views wrong~\cite{Wasscher:1965_a} went largely 
unnoticed and their link to weak ferromagnetism
was not fully understood\cite{Kluczyk:2024_a}.
Even highly cited works~\cite{Shindou:2001_a} about AHE in antiferromagnets
(AFMs) did little to change this state of matters until a prediction
appeared concerning Mn$_3$X systems which are nearly perfectly compensated
on one hand and display large AHE on the other hand.
The prediction~\cite{Chen:2014_a} targeting specifically 
X=Ir (a non-collinear AFM) motivated experimental effort which soon
came to fruition:
observation of AHE in this class of materials, with X=Sn, and theoretical
discussion is reviewed in Ref.~\onlinecite{aa}. In MnTe~\cite{Allen:1977_a}
which is collinear, the AHE was believed by a large part of the
community to be absent until it has
recently been rediscovered in thin layers~\cite{Dominik-PRL} on InP
substrate
independently of previous reports~\cite{Wasscher:1965_a} (and confirmed
for other substrates\cite{SBey} and
in bulk samples~\cite{Kluczyk:2024_a}). Contrary to Mn$_3$X materials,
it is a semiconductor which opens the prospects of changing the material
properties by Fermi level manipulation. It is worth stressing that MnTe
is a special case within the family of other II-VI materials~\cite{Allen:1977_a}
such as MnO or MnS which are cubic~\cite{MG-CA} (and do not break symmetries
which imply\cite{courtesy} vanishing AHE) and, on the other hand, its
NiAs-structure does not support metallic character as in MnAs which is
ferromagnetic moreover.

Effects at zero frequency, such as the AHE, have often their AC counterparts.
Measurements of finite frequency magneto-optical (MO) effect in Mn$_3$X
materials~\cite{Higo:2018_a} and in Mn$_{3}$XN systems~\cite{Johnson}
have once again demonstrated this. Existence of 
magnetic circular dichroism (MCD) at optical range in non-collinear
AFMs has thus been proven (in fact, temperature dependence of MOKE turns 
out to follow closely that of AHE in Mn$_{3}$NiN, for example~\cite{Johnson}).
Spectral shape of MCD in x-ray range (XMCD)
in MnTe has been predicted~\cite{prediction-Hariki} and  
soon confirmed experimentally~\cite{Hariki:2024_a}.  (This effect, in
combination with XMLD, magnetic linear dichroism, has already been used
to visualise antiferromagnetic domains~\cite{Amin:2024_a} in this
material.)
However, changing the applied magnetic field is difficult 
with experiments at synchrotron so we turn our attention to visible-range
MO Kerr effect\cite{Tesarova:2014_a} (MOKE), which requires only tabletop
  experimental setups and allow for continuous sweep of applied
  magnetic field. These effects express themselves as a differential
  absorption of left- and right-circularly polarized light in
  transmission or as a rotation of linear light polarization upon
  reflection. Moreover, despite the advantages of XCMD (in particular
  atom sensitivity), its large inherent line width 
  does not allow for detailed description of the electronic structure around the Fermi level. In this case, optical MOKE measurements are more suitable due to their higher energy resolution.

\begin{figure}
\includegraphics[scale=0.28]{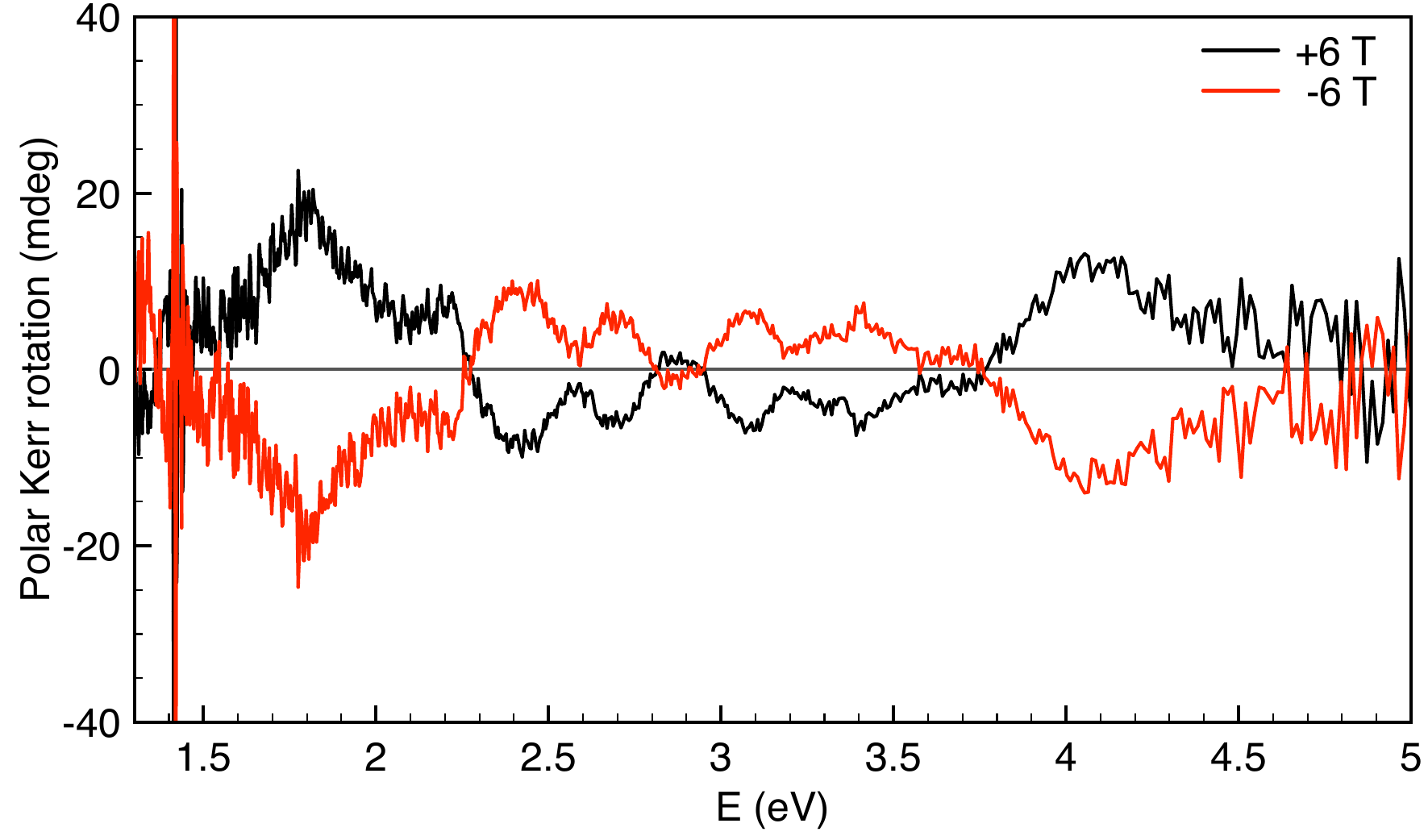}
\caption{Magneto-optical spectra at low temperature (40 K, well below $T_N$)
in strong field ($B=\pm 6$~T, perpendicular to MnTe film) which is
capable of creating substantial imbalance 
  in population between domains of opposite
  polarity~\cite{Hariki:2024_a,Note1}.
  Geometry of the measurement setup is described in the text.}
  \label{fig-01}
\end{figure}

In this work, we present spectra of MOKE in terms of polar Kerr
rotation (for more details see Appendix A) in a collinear antiferromagnet MnTe
for the first time and argue, using comparison to ab initio
calculations, that the origin of the effect is related to mechanism
described in Refs.~\onlinecite{Dominik-PRL,Kluczyk:2024_a} (AFM with
broken $PT$ and $tT$ symmetries~\cite{courtesy}) rather than
to canting of magnetic moments in external magnetic field that generates
magnetisation akin to ferromagnets.

\section{Magneto-optical effects in a collinear antiferromagnet}

\begin{figure}
\begin{tabular}{cc}
  \includegraphics[scale=0.14]{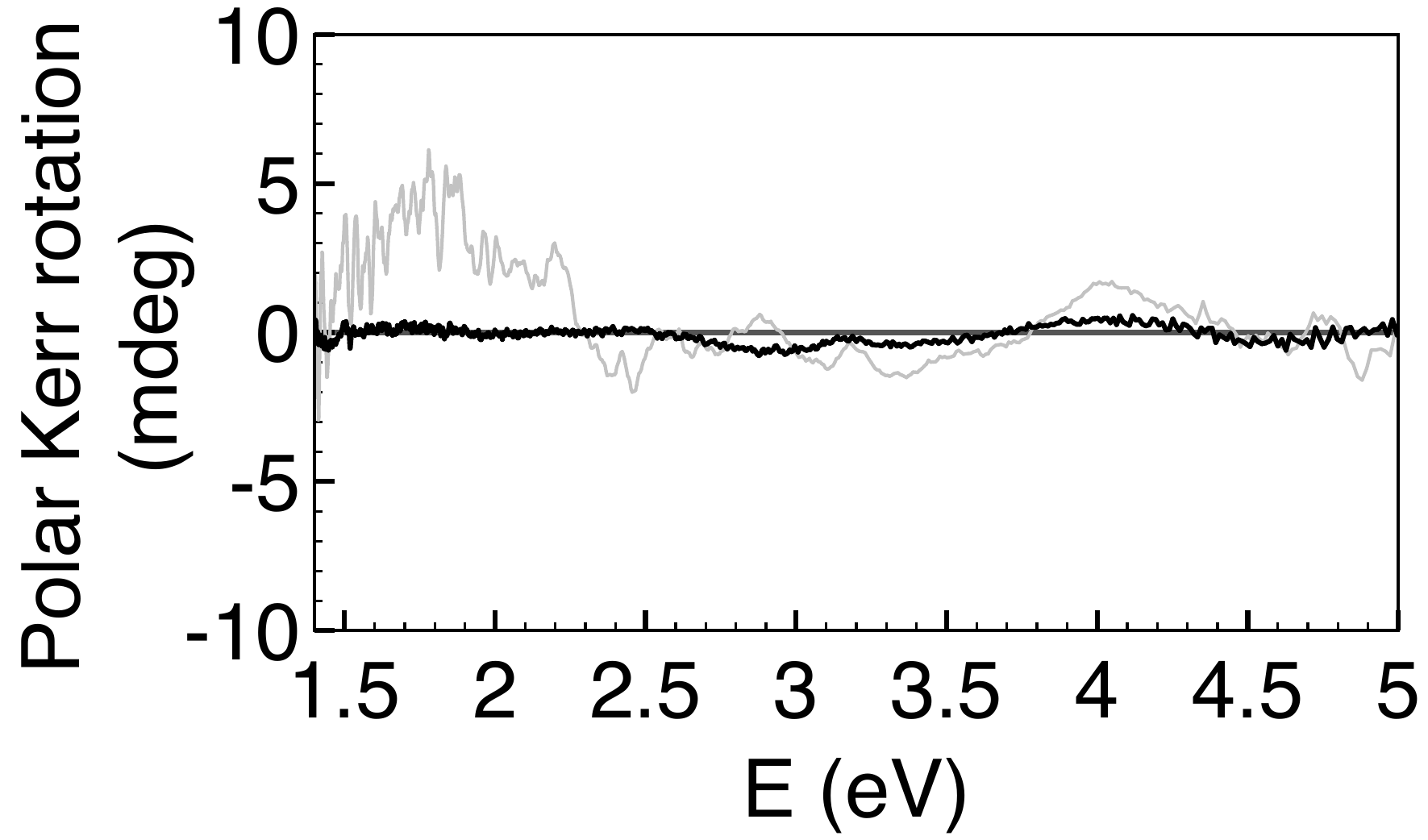} &
  \includegraphics[scale=0.14]{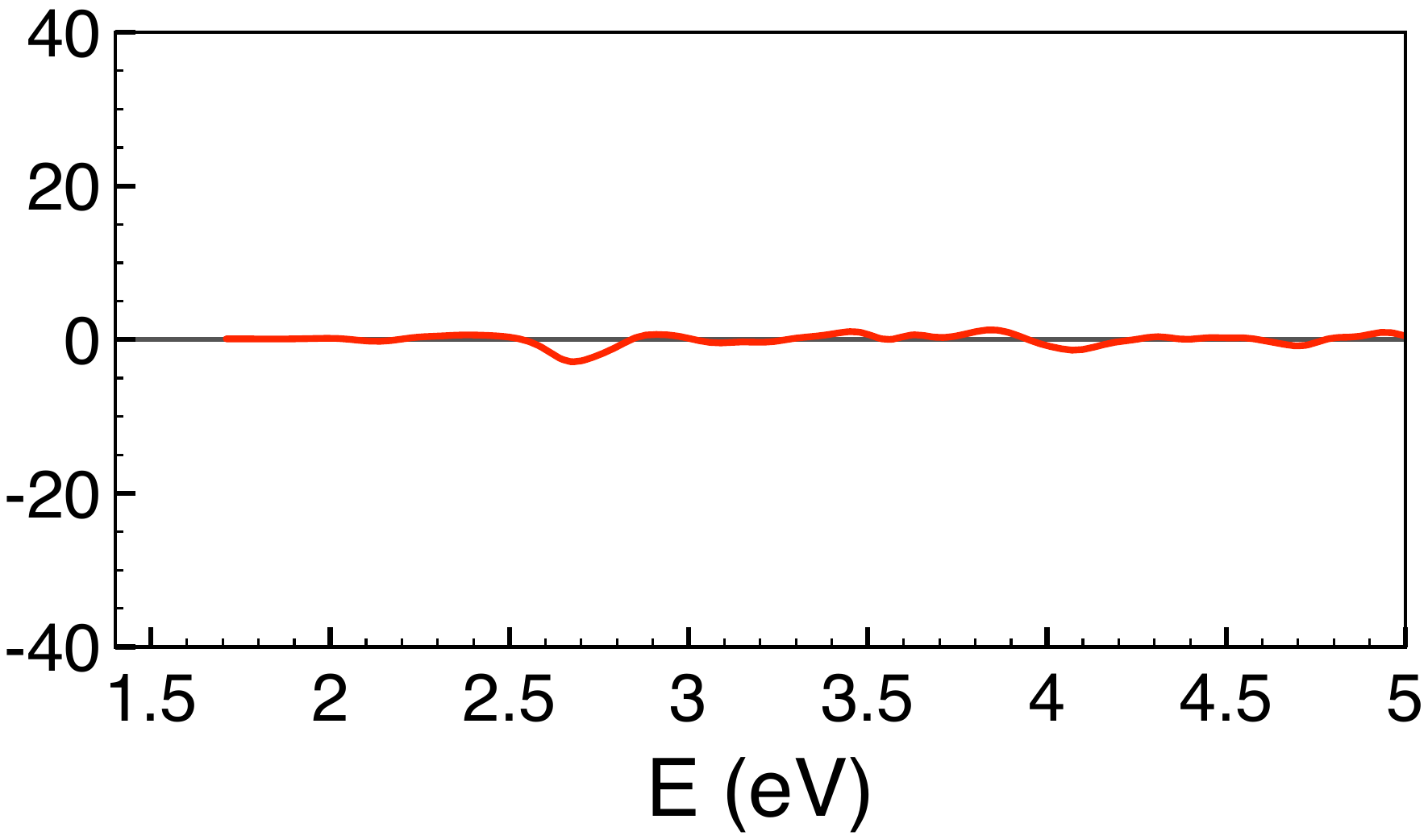} \\  
(a) & (b)
\end{tabular}
\caption{(a) Experimental MOKE spectra in 1~T field above N\'eel temperature (black) and at $T = 40$ K (grey).  
Note the smaller range of vertical axis compared to  Fig.~\ref{fig-01}.
  (b) Calculated MOKE spectra assuming canted moments.}
  \label{fig-02}
\end{figure}

MnTe orders antiferromagnetically~\cite{Kriegner:2017_a} below N\'eel
temperature $T_N=310$~K and its semiconducting character is discussed
in Appendix A. Spectral dependence of polar Kerr rotation at low temperature
($T<T_N$) is shown in Fig.~\ref{fig-01}. The relation of MOKE 
to magnetic order is evidenced by its vanishing above $T_N$, see
Fig.~\ref{fig-02}(a) where spectrum at $T=360$~K (black) is compared to spectrum at $T = 40$ K (grey). These data can be viewed as the
counterpart to XMCD in Fig.~2(a) of Ref.~\onlinecite{Hariki:2024_a} in optical
range and magnetic field (direction and strength) was chosen the same for
the sake of easier comparison. Spectral amplitude of MOKE scales as an
odd function of magnetic field
applied in perpendicular direction to the MnTe thin film; we observe
no clear saturation up to the measurement shown in Fig.~\ref{fig-01}.

Low temperature MOKE spectrum exhibits a positive (for $B=+6$~T)
  polar Kerr rotation at energies below 2.3 eV, followed by negative
  rotation region between 2.3 and 3.9 eV with several negative
  spectroscopic structures located near 2.5, 2.7, 3.1 and 3.4 eV.
  This spectral behavior resembles those reported on other Mn-based
  compounds, such as Mn$_{3}$NiN \cite{Johnson}, Ni-Mn-Ga \cite{Veis}
  or La$_{1/3}$Sr$_{2/3}$MnO$_{3}$ \cite{Mistrik} marking Mn $3d$
  electrons responsible for optical transitions in the visible
  region. Spectral features originate from maxima in joint
  density of states at various points across the Brillouin zone.
Concerning the dependence on the N\'eel vector direction, it has
been shown~\cite{Dominik-PRL,Mazin:2023_a, Betan2024} that upon its in-plane rotation
(measured by angle $\theta$ from the direction where AHE~\cite{Dominik-PRL}
and MOKE are forbidden by symmetry~\cite{courtesy}), the spectra in visible 
range are modulated $\propto \sin 3\theta$. Deviations from this behaviour
are minute, see Fig.~4 in Ref.~\onlinecite{Mazin:2023_a}, except for 
situations where $\sin 3\theta\approx 0$ and the absolute magnitude
of MOKE is small. In our experiments, the in-plane orientation of the
N\'eel vector should be close~\cite{Kriegner:2017_a} to $\sin 3\theta\approx 1$
because of magnetocrystalline anisotropy in the Mn plane (in terms of
crystallography, these planes are perpendicular to [0001], i.e. the $c$-axis).

Regarding microscopic mechanism of the observed MOKE, we must
first consider canting of the Mn moments induced by the applied
magnetic field. At $B=0$, the moments lie nearly perfectly in-plane
(except for the very small canting corresponding likely to arc
seconds~\cite{Kluczyk:2024_a}) and thin layer samples tend to develop
domains~\cite{Amin:2024_a} (we are unaware of any similar
investigation into bulk MnTe). The effect of magnetic field applied
perpendicular to Mn planes (i.e. $\vec{B}\parallel c$ as it
corresponds to $\chi_\perp$ in Ref.~\onlinecite{Kriegner:2017_a})
is then twofold: magnetic moments cant out of the plane and domains of
one polarity~\cite{Note1} favoured~\cite{Dominik-PRL,SBey,Kluczyk:2024_a}  over 
those of the other one. The latter effect is crucial as there would be no
measurable AHE, for example, when the two polarities are equally populated.
On the other hand, the effect of canting has little influence on AHE
but large effect on XMCD (compare the bottom two spectra in Fig.~2(a) of
Ref.~\onlinecite{Hariki:2024_a}). It is therefore important to first
understand what the situation is in MOKE. 
  At this point,
a word of caution is in order. Our interpretation of MOKE (and also the
interpretation of AHE in MnTe~\cite{Dominik-PRL,SBey,Kluczyk:2024_a}) centers
on the idea that the effect can be thought of as a sum of two
contributions: one which depends on $B$ and another which depends on
magnetic order. Specifically, AHE in ferromagnets has often been
written as~\cite{Nagaosa:2010_a} $R_0 B+ R_sM$ and regardless whether
the dependences are simply linear (described by the Hall constant
$R_0$ and an AHE coefficient $R_s$) or more complicated, there is no
room for any 'crossed term' in this scheme. The latter are nevertheless
permitted by Onsager relations.\cite{Zyuzin:2021_a}  Based on the
present experiments, we cannot exclude the possibility of MOKE being
governed by some kind of such product between a variable related to
magnetic order and $B$.

Returning to the usual scheme (where magnetic field plays role only
indirectly through magnetic order), ab initio models of
MOKE spectra (explained in detail in Appendix~B) show that in optical
regime, canted Mn moments cannot possibly explain the signal measured
at $B=6$~T. 
{As shown in Fig.~\ref{fig-02}(a),
the MOKE spectrum measured above $T_N$ exhibits different spectral behavior 
than the spectrum below $T_N$. At energies below 2.5 eV, the spectrum at 
$T>T_N$ is virtually zero. Above 2.5 eV the spectral dependence is 
significantly different from the spectrum at $T<T_N$ and more resembles the 
spectrum in Fig.~\ref{fig-02}(b) which is calculated for magnetic moments
canted by as much as 5 deg.} Canting of few degrees at fields up to 10~T 
can be expected (see p.~4 in Ref.~\onlinecite{Hariki:2024_a}), 
{ but still leads to Kerr rotation 
that is significantly smaller than the measured signal at $T<T_N$.} 
This suggests that the observed effect is related to collinear
order with domains (of opposite polarities) out of balance. There is
an important consequence to this finding: MOKE shown in Fig.~\ref{fig-01}
should effectively only depend on magnetic order through the domain
imbalance.\cite{Note1} In 
other words, as the canting caused by applied magnetic field has only
negligible effect, the increase of signal in stronger fields should
be attributed to gradual suppression of one polarity of
domains~\cite{Kluczyk:2024_a}.

\begin{figure}
\includegraphics[scale=0.5]{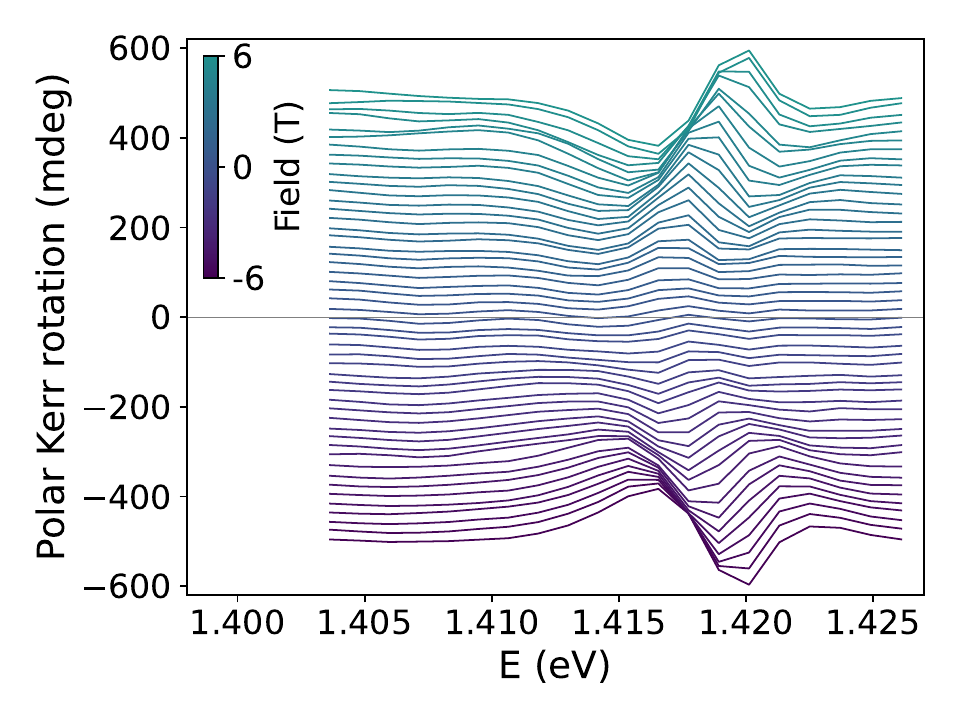}
\caption{Stacked plot of low-temperature MOKE at energies close to $E_g$ of MnTe
  subject to magnetic field $\vec{B}\parallel c$. }
  \label{fig-03}
\end{figure}

We also observe a comparatively large MOKE signal at energies close to
the MnTe band gap energy $E_g$ (in Ref.~\onlinecite{Kriegner:2016_a}, value
$E_g=1.46\pm 0.10$~eV at 40 K was reported). This feature, shown
in Fig.~\ref{fig-03}
becomes stronger and blue-shifts under applied magnetic field (we observe
no similar shift in energy with spectral features far above $E_g$). While we
leave the explanation of this phenomenon for future research, we remark that 
it could either be related to the large sensitivity of MnTe band gap
to rotations of magnetic moments out of the manganese basal
plane~\cite{pfjr:2023} or to magneto-optical activity of the InP substrate.
Accidentally, band gap of InP is very close to $E_g$ and moreover, excitonic
features in optical spectra~\cite{Bimberg:1977_a} blue-shift with applied 
magnetic field. Magneto-optical experiments on bulk MnTe~\cite{Kluczyk:2024_a}
should easily distinguish between these two mechanisms.

\section{Conclusions}

The spectral dependence of 
magneto-optical polar Kerr rotation was
measured at temperature below $T_{N}$ in a collinear antiferromagnet,
which cannot be explained by ferromagnetic moment induced by external
magnetic field. This way, we demonstrated that non-collinear
order~\cite{Higo:2018_a} is not essential for the effect to
appear, in agreement with expectations~\cite{Mazin:2022_a}
associated to the class of magnetic systems called altermagnets (a
more detailed symmetry analysis of MOKE-active collinear systems
has been recently given by Radaelli~\cite{Radaelli:2024_a}).
Moreover, non-vanishing linear magneto-optical effect opens
possibilities for future use of collinear antiferromagnets in
applications where magneto-optical effects are required to be stable
under externally applied magnetic fields.

\begin{figure}
\includegraphics[scale=0.28]{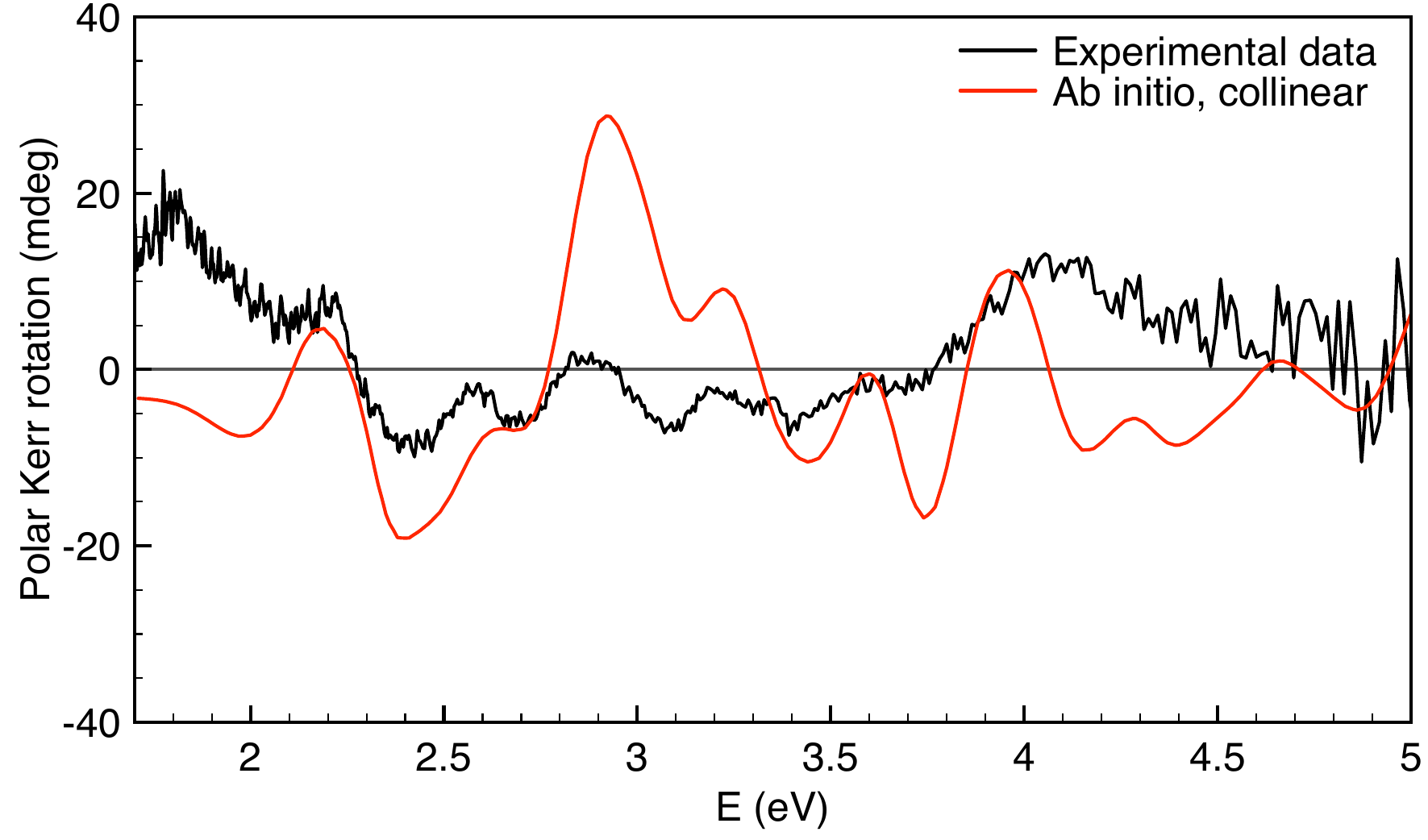}
\caption{Comparison of data from Fig.~\ref{fig-01} to ab initio
  calculations in the collinear state as described in Appendix B.}
  \label{fig-04}
\end{figure}

\section*{Acknowledgments}

Our work benefited from discussions with T. Jungwirth, D. Kriegner and
J. Koloren\v c and we express our gratitude to them as well as to funding
sources from GA \v CR (under contract 22-21974S), Austrian Science
Funds, grant No. AI0656811 and University of Linz, grant
No. LI1113221001.

Part of the experiments were carried out at MGML ({\tt mgml.eu}), which is
supported by the Ministry of Education, Youth and Sports, Czech Republic
within the program of the Czech Research Infrastructures (project no.
LM2023065).

\begin{appendix}

\section{Experimental setup and sample properties}

We used rotating analyzer technique to precisely measure the polar Kerr rotation of the sample. The applied magnetic field pointed in out-of-plane direction with respect to the sample surface and the angle of the light incidence was nearly normal. Our experimental technique comprised a wide spectral laser driven light source (Energetiq EQ-99X-FC) a set of parabolic mirrors to collimate the light beam from the source and focus it on to the sample in the Quantum Design PPMS cryostat with superconducting coil. The light passes two polarizers in the experimental setup. First polarizer defines the linear polarization of light prior to the reflection on the sample and the second polarizer after the sample (called analyzer) is rotating several degrees around the crossed position with the first polarizer. The light is finally detected by CCD fiber spectrometer (Ocean Insight QE Pro).  From the intensity dependence on the analyzer angle, one can calculate the value of the MOKE. 

Our sample was a thin layer (35 nm) of MnTe grown by MBE (molecular
beam epitaxy) on InP, described elsewhere.\cite{Kriegner:2016_a}
In that work, optical absorption edge was confirmed to be close to
$E_g$ and hole density inferred from Hall measurements was
of the order of $10^{19}$ per cubic cm which renders Fermi energy
shift very small~\cite{Kluczyk:2024_a} so that its non-zero
value likely has no measurable effect on MOKE.

\section{Ab initio calculations}

Measured data shown in Fig.~\ref{fig-01} can be compared to models
based on ab initio calculations of electronic structure; the two methods
of choice here are the augmented plane wave (APW) and projector augmented-wave
(PAW) ones. 

Using Kubo formula,~\cite{Tesarova:2014_a}
conductivity tensor $\sigma_{ij}(\omega)$ is first calculated and then
converted to $\epsilon_{ij}$ in a standard way~\cite{Mazin:2023_a}.

Spectral dependence of polar Kerr rotation was
  calculated from the spectral dependence of permittivity tensor
  obtained from ab initio methods. Since the thickness of the layer was only 35 nm we had to consider a model structure consisting the MnTe layer on InP substrate. For this purpose we used 4x4 Yeh's formalism for anisotropic multilayers \cite{Yeh}. 

Unlike in the case of  canted moments, we now argue that measured Kerr rotation
in optical range can be accounted for by density functional theory (DFT)
calculations assuming perfectly collinear order.
In Fig.~\ref{fig-04}, the calculated MOKE spectrum was scaled down by
a factor of 2. This would correspond to domain imbalance of 75~\% (and
this parameter should of course depend~\cite{Note1} on applied magnetic field).

Collinear DFT calculations have been performed using linearised APW
method (implemented in {\tt wien2k}) and data in
Fig.~\ref{fig-02}(b) were generated from non-collinear PAW calculation (in VASP)
with GGA and Hubbard $U=4$~eV (Dudarev scheme\cite{Dudarev:1998_a};
energy cut-off 400~eV). Lattice constants~\cite{Kriegner:2017_a}
$c=0.671$ and $a=0.414$~nm were assumed.

\end{appendix}

\def\urlprefix{}
\def\url#1{}\bibliography{lit}

\end{document}